\documentclass[a4paper,fleqn,usenatbib]{mnras}
\usepackage{color}
\usepackage{graphicx}	       % Including figure files
\usepackage{amsmath}           % Advanced maths commands
\usepackage{amssymb}	       % Extra maths symbols
\usepackage{url}               % For url's, has \url{}, simpler than hyperref
\usepackage{multirow}          % For tables, a table space of several rows.
\usepackage{float}
\usepackage{savesym}
\usepackage{amsmath}
\savesymbol{iint}
\usepackage[varg]{txfonts}
\restoresymbol{TXF}{iint}
\def\apjl{ApJL}
\def\aj{AJ}
\def\apj{ApJ}
\def\pasp{PASP}

\def\apjs{ApJS}
\def\araa{ARAA}
\def\aap{A\&A}
\def\nat{Nature}
\def\mnras{MNRAS}
\newcommand{\amin}{$^{\prime}$}                  % arcminute
\newcommand{\asec}{$^{\prime \prime}$}           % arcsecond
\newcommand{\adeg}{$^{\circ}$}                   % degree

\newcommand{\amindot}[2]{\mbox{#1$\stackrel {\prime}{_{\bf \cdot}}$#2}}        % Fraction of minute
\newcommand{\asecdot}[2]{\mbox{#1$\stackrel {\prime \prime}{_{\bf \cdot}}$#2}} % Fraction of second

\newcommand{\um}{$\mu$m}                         % micrometers
\title[94\,Ceti: a triple star with a planet and dust disc] % Page header
{94\,Ceti: a triple star with a planet and dust disc        % Title page
\thanks{This publication is based on observations with {\it Herschel} which is an ESA space observatory with science instruments provided by European-led Principal Investigator consortia and with important participation from NASA.}$^,$ \thanks{It is also based on data acquired with the Atacama Pathfinder Experiment (APEX). APEX is a collaboration between the Max-Planck-Institut f\"ur Radioastronomie, the European Southern Observatory, and the Onsala Space Observatory.}}

\author[J. Wiegert
    and V. Faramaz
    and F. Cruz-Saenz de Miera
       ]{J. Wiegert,$^{1}$\thanks{E-mail: \url{joachim.wiegert@chalmers.se}}
         V. Faramaz,$^{2}$
         F. Cruz-Saenz de Miera$^{3}$
\\ % Affiliations
$^{1}$Department of Earth and Space Sciences, Chalmers University of Technology, Onsala Space Observatory, SE-439 92 Onsala, Sweden\\
$^{2}$Instituto de Astrof\'{i}sica, Pontificia Universidad Cat\'{o}lica de Chile, Vicu\~{n}a Mackenna 4860, 7820436 Macul, Santiago, Chile\\
$^{3}$Instituto Nacional de Astrof\'{i}sica, Optica y Electr\'{o}nica, Luis Enrique Erro 1, Tonantzintla, Puebla, 72840, M\'{e}xico
}
\date{Accepted 2016 July 11. Received 2016 July 9; in original form 2016 April 26}
\pubyear{2016}
\begin{document}
\label{firstpage}
\pagerange{\pageref{firstpage}--\pageref{lastpage}}
\maketitle

\begin{abstract}
% Context - Can be left empty
94\,Ceti is a triple star system with a circumprimary gas giant planet and far-infrared excess. Such excesses around main sequence stars are likely due to debris discs, and are considered as signposts of planetary systems and, therefore, provide important insights into the configuration and evolution of the planetary system. 
% Aims
Consequently, in order to learn more about the 94\,Ceti system, we aim to precisely model the dust emission to fit its observed SED and to simulate its orbital dynamics.

% Methods
We interpret our APEX bolometric observations and complement them with archived \textit{Spitzer} and \textit{Herschel} bolometric data to explore the stellar excess and to map out background sources in the fields. Dynamical simulations and 3D radiative transfer calculations were used to constrain the debris disc configurations and model the dust emission.

% Results
The best fit dust disc model for 94\,Ceti implies a circumbinary disc around the secondary pair, limited by dynamics to radii smaller than 40\,AU and with a grain size power-law distribution of $\sim a^{-3.5}$. This model exhibits a dust-to-star luminosity ratio of $4.6\pm 0.4\,\times 10^{-6}$. The system is dynamically stable and N-body symplectic simulations results are consistent with semi-analytical equations that describe orbits in binary systems. In the observations we also find tentative evidence of a circumtertiary ring that could be edge-on.
\end{abstract}


% ---------------------------------------------------------------------------- %

\section{Introduction}

Approximately half of the stars belong to binary or multiple systems \citep{duquennoy1991,lada2006,eggleton2008,raghavan2010,duchene2013}. Stellar companions may have a large impact on planetary formation processes. In particular, they are found to truncate circumstellar discs and put a limit on the extent of material available for planetary formation \citep{artymowicz1994,jangcondell2008,andrews2010,jangcondell2015}. They are also expected to stir and increase the eccentricities and relative velocities of planetesimals and planetary embryos that might have formed around the primary \citep{quintana2007,thebault2009,rafikov2015}. Consequently, multiple star systems generate significant collisional activity which may shorten the lifetime of a disc and make it difficult to form planets. Observations tend to support this, since it was found that protoplanetary discs are half as common in young (a few Myr) binary stars as compared to single stars \citep{cieza2009,kraus2012}. In addition, these discs are much less massive compared to those found around single stars \citep[by a factor $\sim 5-25$, ][]{harris2012} and were found to be more short-lived as well \citep{duchene2010}.

However, debris discs have been found around old binary stars, first with \textit{Spitzer} \citep{trilling2007} and then with \textit{Herschel} \citep{rodriguez2015}. They were detected through their host star infrared excess emission, signature of the presence of micron-sized dust grains. These grains are expected to acquire unstable orbits due to both gravitational and non-gravitational perturbations. The latter scenario includes radiation pressure and Poynting-Robertson drag that effectively clear dust clouds of small and large grains ($<1\,$\um\ and between 1\,\um\ to 1\,mm, respectively). As such, dust in circumstellar discs must be continuously replenished from collisional processes in rings of parent bodies (planetesimals) akin to that of the asteroid and Edgeworth-Kuiper belts in our Solar System \citep{artymowicz1997,krivov2008,wyatt2008,moromartin2013}. These underlying reservoirs of large bodies show that the presence of a stellar companion does not necessarily hinder the formation of the building blocks necessary to form planets.

The presence of at least one stellar companion does not necessarily inhibit planetary formation either, since numerous planets were found in binary or multiple systems. However, the majority of these planets were found in wide binary systems \citep{mugrauer2005,daemgen2009,muterspaugh2010,bergfors2013}. In the cases where the companion was not expected to significantly affect planetary formation processes, it is not surprising that these systems were found to be hosts to planets \citep{eggenberger2007,desidera2007,duchene2010} and that the properties of circumstellar protoplanetary and debris discs in binaries are found to be similar to those hosted by single stars for separations larger than several tens of AU \citep{kraus2012,harris2012,rodriguez2015}.

According to the Open Exoplanet Catalogue\footnote{\url{http://www.openexoplanetcatalogue.com/}} (during of the summer of 2016), out of the 130 multiple star systems with planets, only less than one fifth (23) are triple star systems. This is expected considering that triple star systems represent less than one fifth of the multiple star systems \citep{raghavan2010}. In most cases of triple star systems, the planet is orbiting the primary star with a secondary stellar binary surrounding the primary on a wider orbit.

In addition, investigating associated debris discs and tracing the interactions between discs, planets and/or stellar companions can considerably aid our understanding of the dynamical history of a planetary system. This kind of study was carried out for the Solar System with the Nice model of \citet{gomes2005}, but also for the systems of e.g. $\beta\,$Pictoris \citep[see e.g. ][]{dent2014,nesvold2015,millarblanchaer2015,apai2015}, Fomalhaut \citep[see e.g.][]{beust2014,faramaz2015,cataldi2015,lawler2015}, HR\,8799 \citep[see e.g. ][]{moore2013,contro2015,booth2016}, $\tau\,$Ceti \citep{lawler2014}, and HD\,69830 \citep{payne2009}. Therefore, studying in detail close binary or hierarchical systems, where both planets and circumstellar emission have been found, is crucial.

Five systems, out of the 23 found in triple star systems mentioned above, are associated with circumstellar dust emission. These are HD\,178911 \citep{saffe2004,kospal2009}, Fomalhaut \citep{aumann1985,kalas2008,mamajek2013,su2016}, HD\,40979 (e.g. \citealt{kospal2009,dodsonrobinson2011}), 51\,Eridani (e.g. \citealt{rivieremarichalar2014}), and 94\,Ceti (HIP\,14954, HD\,19994). Another possible candidate is L1551 IRS 5, which is a young embedded binary system and potentially a triple star system \citep{lim2006}. Due to the low number of known triple star systems with both planets and dust emission, it is important to study each of these in great detail. We will focus here on the 94\,Cet system.

The third star of 94\,Cet was recently discovered by \citet{roll2011a,roll2012} with astrometry, speckle interferometry, and radial velocity measurements. The system is associated with at least one circumprimary planet \citep{eggenberger2003}, and also shows a far-infrared excess emission \citep{trilling2008,eiroa2013} that possibly originates from circumstellar dust. This is a hierarchical triple star system, at a distance of $22.6 \pm 0.1$\,pc from the Sun, where 94\,Cet\,A is an F8\,V star, and 94\,Cet\,B and C are both M dwarfs that form a binary pair that together orbits the primary on a 2029 year long orbit.

The infrared excess of 94\,Cet was detected by \citet{trilling2008} with \textit{Spitzer}/MIPS at 70\,\um\ and later confirmed, at 100 and 160\,\um\, by \citet{eiroa2013} as part of the key project DUNES (DUst around NEarby Stars) of \textit{Herschel} \citep{pilbratt2010}. Infrared excesses at such long wavelengths originate from the thermal emission of dust particles surrounding the star. To improve the coverage of the spectral energy distribution (SED) we observed this system with APEX (the Atacama Pathfinder EXperiment) at 870\,\um .

\citet{eiroa2013} found that the inferred dust temperature corresponds to a black-body radial distance from the primary star that corresponds to a dynamical unstable region due to the companion stars. Another hint that this emission may not be associated with 94\,Cet\,A is that there is a marginally significant offset between the expected and observed position of the source, if it is co-moving with 94\,Cet\,A. It may thus be a background source or be associated with the other stellar members of this system.

The structure of this paper is as follows; we present the stellar and system properties in Section\,2, and we summarize the observations, data reduction, and the observational results in Section\,3. In Sections\,4 and 5, we discuss the nature of the excess, the extended emission, and background sources. Assuming that the excesses originate from disc(s), we apply both dynamical simulations and radiative transfer simulations and present these results in Section\,6, and summarize our conclusions in Section\,7.

\section{Stellar and system properties}

The binary nature of 94\,Cet was first discovered by Admiral Smyth in 1836 \citep{raghavan2006,smyth1844} who was able to resolve the stars (they share proper motion, $\mu_\alpha = 194.56 \pm 0.37$\,mas\,yr$^{-1}$ and $\mu_\delta = -69.01 \pm 0.30$\,mas\,yr$^{-1}$). The orbital parameters have been constrained and refined by \citet{hale1994}, and more recently by \citet{roberts2011}. The companion star is a binary with two M dwarfs on a one year orbit \citep{roll2011a,roll2012} that together orbits 94\,Cet\,A on a 2029 year long orbit. Fig.\,\ref{hip14954schemorbit} shows a schematic overview of this system and orbital parameters are summarized in Table\,\ref{hip14954orbittable}.

\begin{table}
    \caption{94\,Cet orbital parameters.}
    \label{hip14954orbittable}
    \begin{center}
    \begin{tabular}{lccc}
\hline\hline
                              & Outer                      & Inner                          & Planetary \\
                              & orbit$^a$                  & orbit$^b$                      & orbit$^c$ \\
\hline
    Semi-major                &\multirow{2}{*}{$220 \pm 5$}&\multirow{2}{*}{$0.99 \pm 0.02$}&\multirow{2}{*}{$1.42 \pm 0.01$}\\
    axis (AU)                 &                            &                                &                 \\
    Period (yr)               & $2029 \pm 41$              & $1.04 \pm 0.01$                & $1.47 \pm 0.01$ \\
    Eccentricity              & $0.26 \pm 0.01$            & $0.36 \pm 0.01$                & $0.30 \pm 0.04$ \\
    Inclination to            &\multirow{2}{*}{$104 \pm 2$}&\multirow{2}{*}{$108.5 \pm 0.7$}&\multirow{2}{*}{$\ldots$}\\
    LOS, $i$ (\adeg)          &                            &                                &                 \\
    Arg. of                   &\multirow{2}{*}{$342 \pm 7$}&\multirow{2}{*}{$334.9 \pm 2.3$}&\multirow{2}{*}{$\ldots$}\\
    periapsis $\omega$ (\adeg)&                            &                                &                 \\
    Long. of asc.             &\multirow{2}{*}{$97 \pm 2$} &\multirow{2}{*}{$190.9 \pm 1.0$}&\multirow{2}{*}{$\ldots$}\\
    node $\Omega$ (\adeg)     &                            &                                &                 \\
\hline
    \end{tabular}
    \end{center}
    \begin{list}{}{}
    \item[References.] $^a$ \citet{roberts2011}. $^b$ \citet{roll2011a,roll2012}. $^c$ \citet{mayor2004}
    \end{list}
\end{table}

In the CORALIE survey, that is based on radial velocity measurements, a planetary companion around the primary star was also discovered \citep{queloz2004,mayor2004}, designated 94\,Cet\,Ab. This planet has a mass of $m_b\,\sin\,i = 1.7\,M_{\rm Jup}$, period of 536 days, and a semi-major axis of 1.4\,AU.

The properties of the two companion stars are not well-known as previous works were based on the assumption that 94\,Cet is a binary. However, their masses have been estimated by \citet{roll2011a,roll2012} with radial velocity measurements. These masses suggest them to be M class main sequence stars (the primary star's age is between 1 and 5 Gyr) and so we may use other studies to infer additional properties. From \citet[][pp. 388-389]{cox2000} we find that the masses $0.55\,M_\odot$ and $0.34\,M_\odot$ of main sequence stars correspond to the spectral classes M0 and M3, respectively. \citet{bessell1991} showed that these spectral classes have effective temperatures of 3700\,K and 3300\,K and luminosities of $0.05\,L_\odot$ and 0.01-\,$0.015\,L_\odot$, respectively (see also \citealt{rajpurohit2013}). With these data we approximate their radii (Stefan-Boltzmann's law) to $0.6$ and $0.3\,R_\odot$ and estimate their surface gravity to $\log g\,=\,4.7$ and $5.0$ respectively, (see Table\,\ref{stellarproperties}). 

\begin{table*}
    \caption{Stellar properties.}
    \label{stellarproperties}
    \begin{center}
    \begin{tabular}{lccc}
    \hline\hline
    Names      & 94\,Cet\,A               & 94\,Cet\,B & 94\,Cet\,C \\
               & HIP\,14954\,A, HD\,19994 &            &            \\
\hline
    ICRS (J2000) R.A.       & 3$^{\rm h}$12$^{\rm m}$46\fs 43 &            $\ldots$ & $\ldots$            \\
    ICRS (J2000) Dec.      & -1\adeg 11\amin \asecdot{45}{96} &            $\ldots$ & $\ldots$            \\
    Spectral class                   &   F8\,V - G0\,IV & M0\,V$^{a*}$ & M3\,V$^{a*}$ \\
    Effective temperature (K)             &              6187 &   3700$^{b\dagger}$ &   3300$^{b\dagger}$ \\
    Luminosity ($L_\odot$)                &              3.85 &   0.05$^{b\dagger}$ &   0.01$^{b\dagger}$ \\
    Mass ($M_\odot$)                      &          1.34$^c$ &            0.55$^d$ &            0.34$^d$ \\
    Radius ($R_\odot$)                    &$1.82 \pm 0.07 ^e$ &     0.55$^\ddagger$ &     0.31$^\ddagger$ \\
    $\log (g)$                            &              4.24 &        4.7$^\sharp$ &       5.0$^\sharp$  \\
    Metallicity [Fe/H]                    &              0.21 &     $\ldots$        &            $\ldots$ \\
    Age (Gyr)$^\amalg$                    &       1.17 - 5.67 &         1.17 - 5.67 &         1.17 - 5.67 \\
    Distance (pc)                         &    $22.6 \pm 0.1$ &      $22.6 \pm 0.1$ &      $22.6 \pm 0.1$ \\
    Proper motion (mas yr$^{-1}$), R.A.   &   $195^{f\Delta}$ &    $195^{f\Delta}$  &            $\ldots$ \\
    Proper motion (mas yr$^{-1}$), Dec.   &   $-67^{f\Delta}$ &    $-67^{f\Delta}$  &            $\ldots$ \\
\hline
    \end{tabular}
    \end{center}
    \begin{list}{}{}
    \item[{\bf Notes.}] General reference is \citet[][and references therein]{eiroa2013} except when stated otherwise.$^*$ Estimate from tables \citep[][pp. 388-389]{cox2000} using the known dynamical mass.$^\dagger$ Estimate from tables \citep{bessell1991} using the spectral class.$^\ddagger$ Estimate from Stefan-Boltzmann law $L \approx 4 \pi R^2 \sigma T^4$.$^\sharp$ Approximated from radius and mass to find a spectral model in a PHOENIX grid.$^\amalg$ Age range based on both X-ray luminosities and activity index $\log R'_{\rm HK}$ \citep{eiroa2013}.$^\Delta$ 94\,Cet's system proper motion is $\mu_\alpha = 194.56 \pm 0.37$\,mas\,yr$^{-1}$ and $\mu_\delta = -69.01 \pm 0.30$\,mas\,yr$^{-1}$ (\citealt{eiroa2013} and SIMBAD Astronomical Database).
    \item[{\bf References.}]$^a$ \citet[][pp. 388-389]{cox2000}.$^b$ \citet{bessell1991}.$^c$ \citet{mayor2004}.$^d$ \citet{roll2011a,roll2012}.$^e$ \citet{fuhrmann2008}.$^f$ SIMBAD Astronomical Database \url{http://simbad.u-strasbg.fr/simbad/}.
    \end{list}
\end{table*}

\section{Observational data}

In this work we made use of photometric data to obtain properties of the dust conforming the disc surrounding 94\,Cet, and of imaging data used to make a spatial analysis of the whole system. In this section we explain how we obtained and processed the \textit{Herschel} and APEX-LABOCA data plus we present the supplementary data used in our analysis. The complete observational data are presented in Table \ref{obslog}.

\begin{figure*}
    \begin{center}
        \includegraphics[width=157mm]{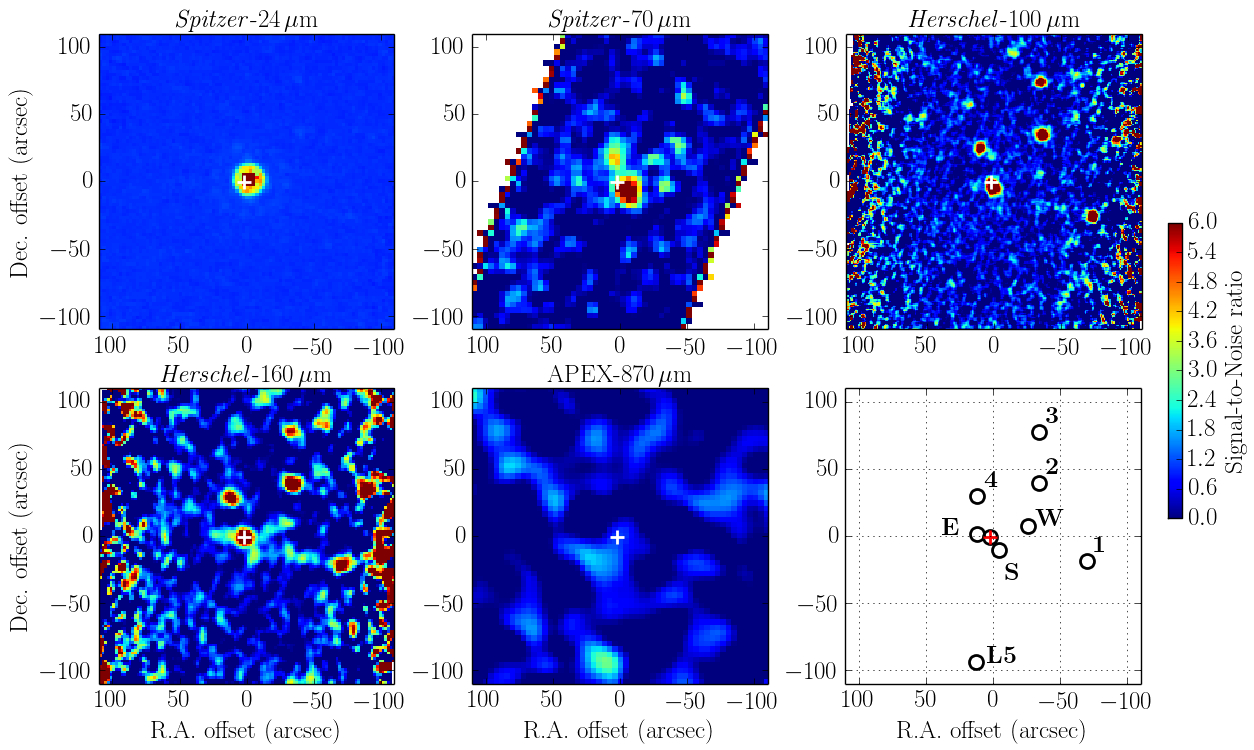}
        \caption{SNR maps of the \textit{Spitzer}/MIPS 24 and 70\,\um\ images (as extracted from the \textit{Spitzer} archive, \citealt{trilling2008}), the \textit{Herschel}/PACS 100 and 160\,\um\ images (observed by \citealt{eiroa2013}), and our APEX-LABOCA 870\,\um\ observation. The lower right image identifies background sources found in the field, numbered in the same order as in Tables\,\ref{spectralindices} and \ref{labocafluxes}. The stellar position is indicated with a cross, and E, S, and W denote the eastern, southern, and western extensions (see the text).}
        \label{allfields}
    \end{center}
\end{figure*}

\begin{table*}
    \caption{Observation log of data used in our analysis.}
    \label{obslog}
    \begin{center}
    \begin{tabular}{llcclrlc}
    \hline\hline
Instrument/ & Obs/Pgm ID/ & $\lambda_{\rm eff}$ & Beam         & Observing Date  & $t_{\rm int}^a$ & Field centre        & Offset$^b$\\
       Mode & AORKEY      &  (\um)              & FWHM (\asec) & year-mo-day     & (s)             & coordinates (J2000) &  (\asec) \\
\hline
{\it Spitzer}/MIPS  & 4080640       &  24 &  5.9  & 2005-01-23    &    48 & 3$^{\rm h}$12$^{\rm m}$46\fs 43, $-1$\adeg 11\amin \asecdot{48}{2}  & 2.0 \\ % 2005.06
                    & 4080640       &  70 &  19   & 2005-01-23    &   440 & 3$^{\rm h}$12$^{\rm m}$46\fs 51, $-1$\adeg 11\amin \asecdot{46}{4} & 1.8 \\ % 2005.06
{\it Herschel}/PACS & 1342216129    & 100 &  6.8  & 2011-03-06    &  1440 & 3$^{\rm h}$12$^{\rm m}$46\fs 36, $-1$\adeg 11\amin \asecdot{47}{5} & 5.0 \\ % 2011.18
                    & 1342216130    & 160 & 11.4  & 2011-03-06    &  1440 & 3$^{\rm h}$12$^{\rm m}$46\fs 36, $-1$\adeg 11\amin \asecdot{47}{5} & 6.2 \\ % 2011.18
APEX-LABOCA         & 090.F-9302(A) & 870 & 19.5  & 2012-08-15 -- & 28905 & 3$^{\rm h}$12$^{\rm m}$46\fs 44, $-1$\adeg 11\amin \asecdot{46}{0} &$\ldots$\\% 2012.62
                    &               &     &       & 2012-11-27    &       &   &     \\ % 2012.91
\hline
    \end{tabular}
    \end{center}
    \begin{list}{}{}
    \item[$^a$] $t_{\rm int}$ is the on-source integration time.
    \item[$^b$] Offset between measured main source position and source tabulated coordinates (J2000) of the primary star compensated for proper motion, average offset for PACS is \asecdot{2}{4} \citep{sanchezportal2014} at this observing period (OD 661).
    \end{list}
\end{table*}

\subsection{\textit{Herschel}}

The \textit{Herschel}/PACS \citep{poglitsch2010} observations were taken as part of the DUNES Open Time Key Programme and were published by \citet{eiroa2013}\footnote{The DUNES data archive can be found at \url{http://sdc.cab.inta-csic.es/dunes/jsp/masterTableForm.jsp}}, where a more in-detail description is given (see also \citealt{montesinos2016} for further details). Scan map observations by \textit{Herschel}/PACS were taken with the 100/160 channel combination during OD 661. The \textit{Herschel} observations are summarized here in Table\,\ref{obslog}.

\subsubsection{Data reduction}

The observations were reduced using the Herschel Interactive Processing Environment, {\sc hipe} \citep{ott2010}, version 13.0.0, using the PACS calibration version 69. PACS images are affected by shot noise (frequency dependent, $1/f$), consequently, we used a high-pass filter to reduce these effects. We chose radii of 20 and 25 frames for 100 and 160\,\um , respectively, allowing us to eliminate background structures larger than 82 and 102\,arcseconds. By the nature of the procedure, high-pass filtering results in a flux loss which we rectify by masking all pixels ten times brighter than the standard deviation of non-zero flux pixels. Deglitching was done by using the second level spatial deglitching task in {\sc hipe}. PACS observations were carried out at two different observational angles to decrease striping effects. In the final maps we combined individual scans for each band to increase the signal-to-noise ratio (SNR). Finally we used a drizzling method to obtain a final image scale of 1$\arcsec$ per pixel at 100\,\um\ and 2$\arcsec$ per pixel at 160\,\um\ compared to the native instrument pixel sizes of \asecdot{3}{2} for 100\,\um\ and \asecdot{6}{4} for 160\,\um . 

\subsubsection{Aperture photometry}

Aperture photometry was performed using radii of 5$\arcsec$ and 8$\arcsec$ for 100\,\um\ and 160\,\um , respectively; these particular radii were used because they were found to maximise the SNR by \citet{eiroa2013}. We performed aperture and colour corrections following \citet{balog2014}. To carry out the aperture corrections we divided the integrated flux by 0.521 for 100\,\um\ and by 0.527 for 160\,\um . The colour corrections were done by dividing the aperture corrected flux by 1.033 for 100 \um\ and 1.074 for 160 \um , these values are correct for black body temperatures around 5000\,K (Table\,1 of \textit{Herschel} documentation PICC-ME-TN-038, 2011).

\begin{figure}
    \begin{center}
        \includegraphics[width=80mm]{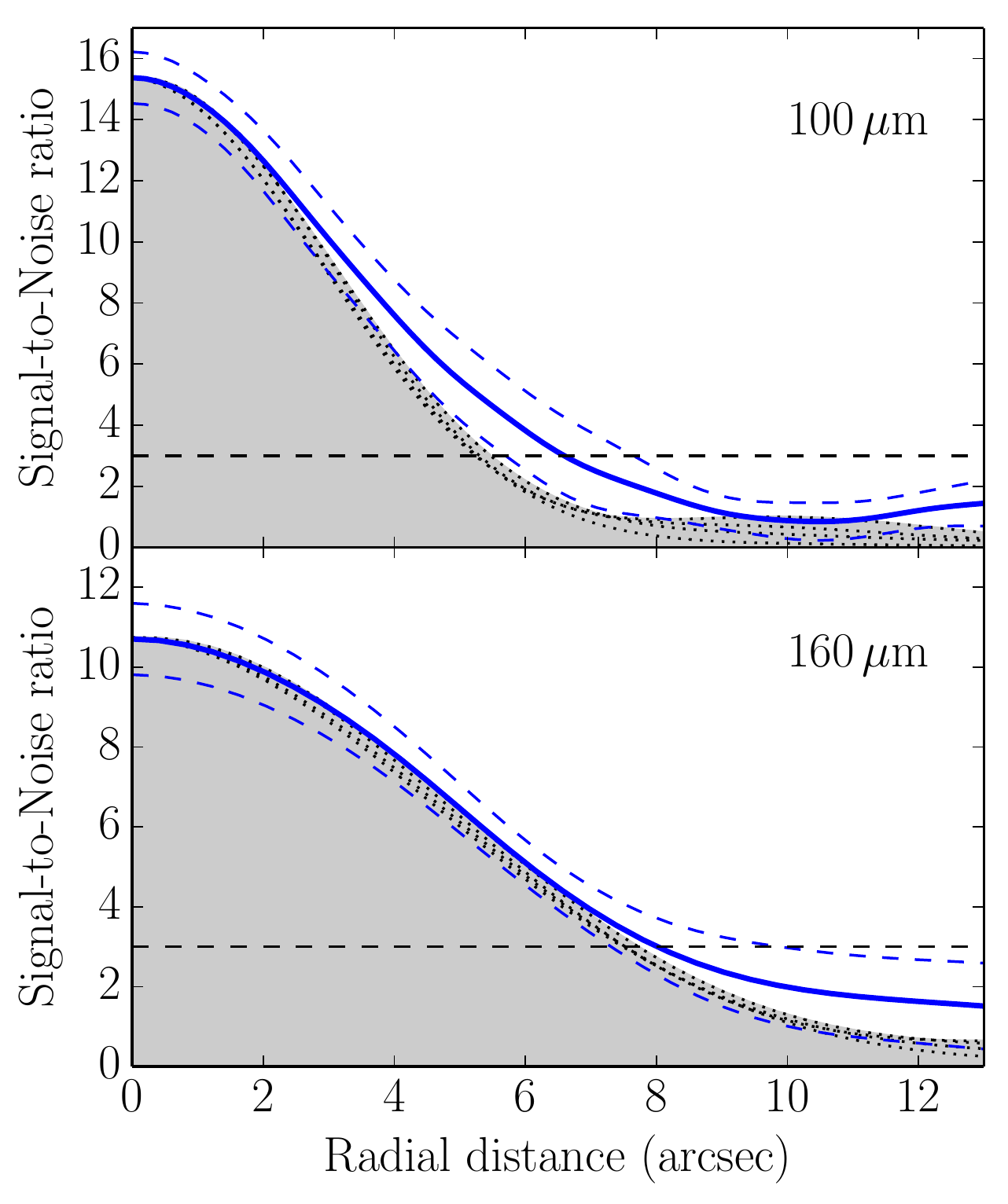}
\caption{In blue, we show the average radial profiles of the central source (continuous) and its uncertainties (dashed) compared with profiles of the $\alpha$\,Bo{\"o} PSF plotted as gray fields and dotted black curves. The $3 \times$\,RMS is indicated with the horisontal dashed black line.}   
        \label{radialprofiles}
    \end{center}
\end{figure}

To check whether or not our source is extended we compared the full width half maximum (FWHM) of 2D Gaussian fits with the beam FWHM. We found that only the central source was marginally extended as we can see in the radial profiles shown in Fig.\,\ref{radialprofiles}. The profile error in this plot are estimated from the standard deviation of the radial profiles to the east, west, south, and north of the central source, and from the observed uncertainty. FWHM of Gaussians fitted on the central source was \asecdot{8}{2} and \asecdot{10}{9} at 100 and 160\,\um , respectively. These correspond to 185 and 246\,AU at the system's distance. The point spread function (PSF) FWHM is \asecdot{6}{8} at 100\,\um\ and \asecdot{11}{4} at 160\,\um\ with a scan speed of 20\asec\,sec$^{-1}$. Thus we used an aperture with a radius of 6$\arcsec$ at 100\,\um , and the aperture correction used on the central source was 0.595 (see also \citealt{marshall2014} for details).

The integrated flux uncertainty was estimated in a process similar to the one explained in \citet{eiroa2013}. We began by measuring the fluxes inside of forty squared apertures; each one of the same area as the aperture used on the 94\,Cet. They were put at random positions, in annulii between 10 to 20\arcsec\ from the source aperture, making sure to avoid all of the nearby sources. Subsequently, the standard deviation of all of these integrated fluxes was multiplied by $\sqrt{1 + 1/N}$, where $N$ is the number of background squares. Lastly, for the final error estimate, we incorporated the calibration error \citep[5\,per\,cent,][]{balog2014} with a quadratic sum.

\subsection{APEX-LABOCA}

APEX\footnote{\url{http://www.apex-telescope.org/telescope/}} is a 12-m millimetre/submillimetre telescope located at 5105\,m altitude on the Llano de Chajnantor, Chile, with a pointing accuracy of 2\,arcsec.

We used the bolometer camera LABOCA (Large APEX BOlometer CAmera, \citealt{siringo2009}) to observe 94\,Cet, project id 090.F-9302(A). The LABOCA operating wavelength is 870\,\um , centred on a 150\,\um\ wide window. The 295 bolometers yield a circular field of view of \amindot{11}{4} and the beam has a FWHM of \asecdot{19}{5}. It uses different mapping modes to fill the undersampled parts of the array; we used the on-the-fly (OTF) mode, in which the maps are scanned back-and-forth linearly row-by-row or column-by-column. Skydips determined the atmospheric opacity with $\tau$ values between 0.19 and 0.40, with an average of $\sim 0.30$ and a precipitable water vapour of PWV $<\,1$\,mm.

The observations were centred on the J2000 coordinates R.A. 3$^{\rm h}$12$^{\rm m}$46\fs 44 and Dec. $-1$\adeg 11\amin \asecdot{46}{0} and produced a \amindot{11}{4}$\times$\amindot{12}{7} sized map. Calibrations yield $5.9 \pm 0.6$\,Jy\,beam$^{-1}$\,$\mu$V$^{-1}$ with Uranus and Neptune as calibrators\footnote{Primary calibrators are Mars, Uranus, and Neptune, and secondary calibrators are listed at \url{http://www.apex-telescope.org/bolometer/laboca/calibration/}}. The data were reduced with {\sc crush}\footnote{Comprehensive Reduction Utility for {\sc sharc-2} \url{http://www.submm.caltech.edu/~sharc/crush/}} using the deep field setting.

The final reduced image has a pixel resolution of 4\asec\,pixel$^{-1}$ and the flux density is given in Jy\,beam$^{-1}$. Point sources' flux densities are measured directly from their peak pixel, and the error estimate is given by the RMS of the background around each source, i.e. in a region extending between two beam radii to 60\asec\ from the source.

\subsection{Ancillary Data}

We took optical data from two different sources Str\"omgren \textit{uvby} photometry from \citet{hauck1997} and Johnson \textit{BV} plus Cousins \textit{I} from \citet{perryman1997}. $JHK_s$ 2MASS data was taken from \citet{cutri2003}. For the mid-infrared regime we took data from the Wide-field Infrared Survey Explorer (\textit{WISE}) and the \textit{Akari} satellites; the \textit{WISE} was taken from the All-Sky data Release Catalogue \citep{wright2010} and the \textit{Akari} data was taken from the \textit{Akari}/IRC mid-IR all-sky Survey \citep{ishihara2010}. We took data from the Infrared Astronomical Satellite \textit{IRAS} from the \textit{IRAS} Point Source Catalogue \citep{helou1988}. Finally, we complemented the SED using \textit{Spitzer} data. \textit{Spitzer}/MIPS \citep{rieke2004} observations are published and summarized in more detail by \citet{trilling2008}. A \textit{Spitzer}/IRS \citep{houck2004} spectrum (PID 102, PI: Werner, \citealt{rebull2008}) extracted from the DUNES database, and not reprocessed by us, is also shown with the SED.

\subsection{Observational results}

Since the angular distance between 94\,Cet\,A and BC during the observations was \asecdot{2}{4} and the 100\,\um\ beam FWHM is \asecdot{6}{8}, the photometry presented here includes the fluxes from all 94\,Cet components.

\begin{figure}
    \begin{center}
        \includegraphics[width=85mm]{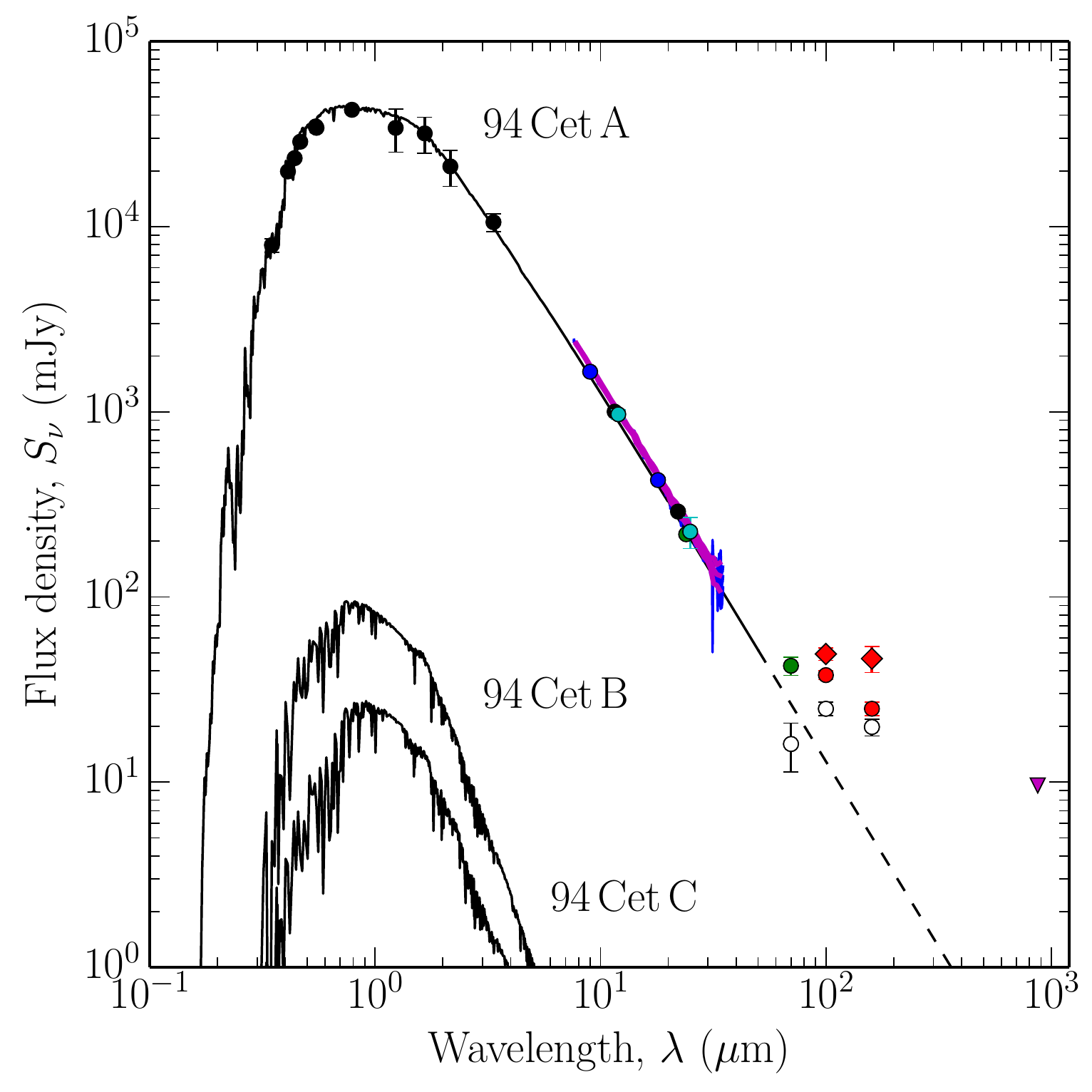}
        \caption{Observed photometry and photospheric model SEDs of 94\,Cet\,A, B, and C. The model photospheres are shown as black lines, the dashed line is an extrapolated Rayleigh-Jeans tail of the stellar photosphere from 50\,\um , under the assumption of blackbody emission. The plotted photometric data points and $1\,\sigma$ uncertainties are also presented in Table\,\ref{allfluxes} (several error bars are smaller than the data point symbol) and the upper limit (triangle) is a $3\,\sigma$ limit. Black data points are \textit{Hipparcos}, 2MASS, \textit{WISE}, and Str\"omgren data, blue are \textit{Akari}, green are \textit{Spitzer}/MIPS, red are \textit{Herschel}/PACS, and magenta upper limit is from APEX-LABOCA. The magenta and blue spectra denote the \textit{Spitzer}/IRS spectrum, where the magenta spectrum is binned to decrease the noise at 30\,\um . The red diamonds represent the combined flux density of the extended emission and the central source, and the empty circles are the excess data (observations subtracted by the photospheric model).}
        \label{seddata}
    \end{center}
\end{figure}

The photometric data is presented in Table\,\ref{allfluxes} and the stellar SED is presented in Fig.\,\ref{seddata}. Excesses are clearly detected, both at 100 and 160\,\um\ (12\,$\sigma$ and 10\,$\sigma$, respectively) and marginally at 70\,\um\ (3.4\,$\sigma$, where $\sigma$ denotes the error estimate at each wavelength). The photospheric spectrum was extracted from the high-resolution PHOENIX/GAIA grid \citep{brott2005} by \citet{eiroa2013} with the parameters summarized in Table\,\ref{stellarproperties}. For the companion stars we used synthetic spectra from the PHOENIX library by \citet{husser2013} (assuming similar metallicities as 94\,Cet\,A, [Fe/H]$\,=\,$0.2). We approximated the long wavelength part of the spectra ($> 10\,$\um\ for B and C, $> 50\,$\um\ for A) with Rayleigh-Jeans tails. The spectra were normalised using the V-band magnitude 11.5 for 94\,Cet\,B$+$C and the B and C flux density ratio of 0.29 \citep{roll2011a,roll2012} was used.

Our PACS photometry results are similar to those of \citet{eiroa2013} (colour corrected; $38.19 \pm 1.75$\,mJy at 100\,\um , and $29.56 \pm 1.16$\,mJy at 160\,\um ), this is caused by the differences in {\sc hipe} versions, in PACS calibration trees, and in aperture corrections. Our 100\,\um\ image has a RMS between 2.0 -- 2.4\,mJy\,beam$^{-1}$, our 160\,\um\ image has a RMS between 3.3 -- 4.0\,mJy\,beam$^{-1}$, and they contain some background sources labeled as numbers 1-5 (1-4 are visible in Fig.\,\ref{allfields}).

\begin{table}
    \caption{Summary of photometry used in the analysis of 94\,Cet.}
    \label{allfluxes}
    \begin{center}
    \begin{tabular}{llc}
\hline\hline
    $\lambda_{\rm eff}$ & $S_{\nu}$ & Photometry \\
    (\um)               & (mJy)     & Reference  \\
\hline
   0.349 & $ 7943    \pm  673 $     & Str\"omgren $u$ (1) \\
   0.411 & $19850    \pm  819 $     & Str\"omgren $v$ (1) \\
   0.440 & $23410    \pm  431 $     & Johnson $B$ (2) \\
   0.466 & $28710    \pm  530 $     & Str\"omgren $b$ (1) \\
   0.546 & $34720    \pm  640 $     & Str\"omgren $y$ (1) \\
   0.550 & $34130    \pm  629 $     & Johnson $V$ (2)\\
   0.790 & $42710    \pm  787 $     & Cousins $I$ (2)\\
   1.235 & $34110    \pm 8895 $     & 2MASS $J$ (3) \\
   1.662 & $31850    \pm 7038 $     & 2MASS $H$ (3) \\
   2.159 & $21120    \pm 4628 $     & 2MASS $K_s$ (3) \\
   3.353 & $10560    \pm 1159 $     & \textit{WISE} (W1) (4) \\
   9     & $ 1645    \pm 19   $     & \textit{Akari} (5) \\
  11.561 & $ 1001    \pm   12 $     & \textit{WISE} (W3) (4) \\
  12     & $  968.8  \pm 58.1 $     & IRAS (6) \\
  18     & $  427.6  \pm 14.2 $     & \textit{Akari} (5) \\
  22.088 & $  289.1  \pm  5.3 $     & \textit{WISE} (W4) (4) \\
  24     & $  218.0  \pm  4.4 $     & MIPS (7) \\
  25     & $  225.3  \pm 42.8 $     & IRAS (6) \\
  70     & $   42.50 \pm  4.76$     & MIPS (7) \\
 100     & $   37.85 \pm  2.08$     & PACS (8) \\
 160     & $   24.92 \pm  2.08$     & PACS (8) \\
 100     & $ 49.20 \pm 3.81^{\ast}$ & PACS (8) \\
 160     & $ 46.46 \pm 7.41^{\ast}$ & PACS (8) \\
 100     & $   38.19 \pm  1.75$     & PACS (9) \\
 160     & $   29.56 \pm  1.16$     & PACS (9) \\
 870     & $\le 9.59\ (3\,\sigma)$  & LABOCA (10) \\
\hline
    \end{tabular}
    \end{center}
    \begin{list}{}{}
    \item[{\bf Notes.}]
$^{\ast}$ Total flux density of both central source and extended emission.
    \item[{\bf References.}]
(1) \citet{hauck1997}.
(2) \textit{Hipparcos} \citep{perryman1997}.
(3) 2MASS Point Source Catalogue \citep{cutri2003}.
(4) \textit{WISE} All-Sky data Release Catalogue \citep{wright2010}.
(5) \textit{Akari}/IRC mid-IR all-sky Survey, II297 in Vizier, colour corrected \citep{ishihara2010}.
(6) \textit{IRAS} Point Source Catalogue, II/125 in Vizier, colour corrected \citep{helou1988}.
(7) \textit{Spitzer}/MIPS, \citet{trilling2008}.
(8) \textit{Herschel}/PACS, this work, colour corrected.
(9) \textit{Herschel}/PACS, estimates by \citet{eiroa2013}, colour corrected.
(10) APEX-LABOCA, this work.
    \end{list}
\end{table}

The emission that appears around the central source at 100 and 160\,\um\ may be associated with 94\,Cet. We estimate the total flux density of the extended source by separating it into an eastern, southern, and western regions designated simply as E, S, and W in Fig.\,\ref{allfields}. The angular distances, from the observed stellar position, for each of these regions are \asecdot{18}{0} for E, \asecdot{17}{1} for S and \asecdot{35}{1} for W; these correspond to projected distances from the main source of 407, 387 and 794\,AU, respectively. The eastern and western extensions are aligned along the projected plane of the companion star's orbit around the primary. The southern extension is along the direction to the companion stars.

To estimate the flux densities for each region, we began by subtracting a standard PACS PSF from the main source. This PSF was normalised to the flux density of the nearby point source 94\,Cet\,2 to avoid underestimating the fluxes by subtracting the extended central emission. The flux densities for each region were then estimated as point sources and the results are presented in Table\,\ref{extensionflux}. The total flux density of the whole 94\,Cet system is listed in Table\,\ref{allfluxes} and plotted in Fig.\,\ref{seddata} as red diamonds.

Among the standard PACS PSFs listed in the \textit{Herschel} technical document PICC-ME-TN-033 (2015), we chose $\alpha$\,Bo{\"o}tis because it has the highest SNR. Additionally, \citet{kennedy2012b} found that the PSF widths varies about 2 -- 4\,per\,cent at 100\,\um\ and $\sim$1\,per\,cent at 160\,\um\ so the effect of chosing another reference star is negligible.

\begin{figure*}
    \begin{center}
        \includegraphics[width=132mm]{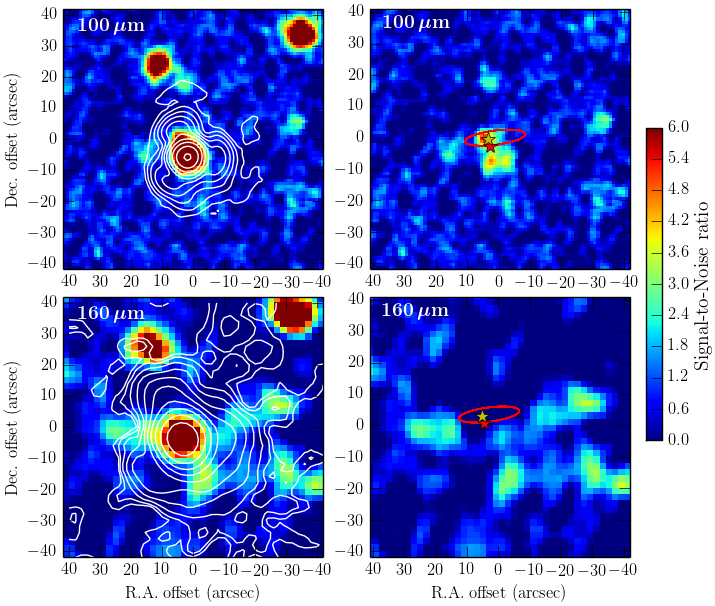}
        \caption{The surrounding area around 94\,Cet. Each row is for each PACS wavelength. The first column shows the observations and the contours of the PSF (white) correctly rotated and positioned on the central source. The PSF is shown in logarithmic scale to emphasize the shape and the positions of the tri-lobes. The second column shows the observations when subtracted by PSFs positioned on the sources. The central source was subtracted by a PSF normalised to the flux density of 94\,Cet\,2. The positions of 94\,Cet\,A (yellow star), 94\,Cet\,BC (red star), and their orbit (red ellipse) around the primary (based on the parameters in \citealt{roberts2011}) are also indicated. 1\,arcsec corresponds to $\sim 23$\,AU.}
        \label{hip14954zoomedandpsf}
    \end{center}
\end{figure*}

\begin{table}
    \caption{Flux densities of source extensions.}
    \label{extensionflux}
    \begin{center}
    \begin{tabular}{rll}
\hline\hline
    Source & $S_{\nu}(100$\,\um $)$ & $S_{\nu}(160$\,\um $)$ \\
           & (mJy)                  & (mJy)                  \\
\hline
    Eastern  & $4.2 \pm 0.5$ & $8.5 \pm 1.8$ \\
    Southern & $2.7 \pm 0.5$ & $5.5 \pm 1.2$ \\
    Western  & $4.5 \pm 0.7$ & $7.6 \pm 2.3$ \\
\hline
    \end{tabular}
    \end{center}
\end{table}

\section{Surrounding sources}

It is imperative that we first distinguish which sources are not part of the 94\,Cet system and which  are likely part of it. We then focus on the latter and attempt to provide first constraints on their positions in the system, i.e., if they are due to circumstellar, circumbinary, or even circumtertiary dust emission.

\subsection{Background contamination}

Background confusion is an important issue at 100 and 160\,\um. When studying cold circumstellar dust, 100\,\um\ is the preferred wavelength because of its sensitivity to the emission of dust with temperatures of $\sim$100\,K and because of the high contrast between the dust and the host star. However, dust emission from high-redshift galaxies may also have significant flux densities at this wavelength which could be misinterpreted as part of the disc emission.

Hence we can estimate the probability of contamination by background galaxies. In figure 7 of \citet{berta2011} it is shown that the number of galaxies with a flux density greater than 2\,mJy beam$^{-1}$ is about $2$ -- $6 \times 10^3$\,deg$^{-2}$, or $1$ -- $3 \times 10^{-2}$\,beam$^{-1}$. This number decreases significantly with increased flux. For example, at 6\,mJy\,beam$^{-1}$ (3 times the background RMS) the number of expected galaxies decreases to $3$ -- $7 \times 10^2$\,deg$^{-2}$, or $1$ -- $3 \times 10^{-3}$\,beam$^{-1}$. In the case of the 94 Cet excess at 100\,\um\ (25\,mJy) the number count has dropped off to $1$ -- $3 \times 10^{-4}$ beam$^{-1}$; this gives the probability of up to 2.3\,per\,cent of coincidental alignment for a given source at these flux levels. As comparison, our estimated probability is lower than the one reported in \citet{krivov2013}, in which they calculated a 4.8\,per\,cent probability of confusion between background galaxies and the coldest debris discs yet found.

As mentioned before, in the PACS image we see a number of background sources (see Appendix A). Those with SNR\,$> 3$ are listed in Table\,\ref{spectralindices} and those in the LABOCA field are listed in Table\,\ref{labocafluxes}. Most of the LABOCA sources are positioned outside the PACS field and not visible in Fig.\,\ref{allfields}.

\subsection{Pointing accuracy}

The average \textit{Herschel} pointing offset ($\sigma_{\rm point}$) for this observing period (OD 661) is \asecdot{2}{36} \citep{sanchezportal2014}. The last column in Table\,\ref{obslog} refers to the observed offset, i.e. the difference between the observed position and J2000 position compensated for 94\,Cet's system proper motion. 

The observed offsets of 94 Cet are $2.1 \sigma_{\rm point}$ and $2.6 \sigma_{\rm point}$ at 100 and 160\,\um , respectively, and while they are marginally significant, there are three scenarios which could explain them. First, the observed emission does not belong to the 94\,Cet system and actually originates from a background galaxy but, as we showed earlier, the probability of this is lower than 3\,per\,cent. Second, the \textit{Herschel} pointing offset, $\sigma_{\rm point}$, could have been larger for these observations, implying that our observed offsets are smaller and the observed emission actually matches the position of the primary star of the system. Third, the observed emission is actually on the expected position for the BC companion pair and originates from a circumbinary debris disc around them.

The second scenario is weakened by the analysis done in \citet{eiroa2013}, in which they found that dust emission temperature corresponds to a distance to the primary star where no stable orbits can exist due to influences from the secondary. Meanwhile, the third scenario is strengthened by the orbital measurements from \citet{roberts2011}, which locate the BC companion pair at observed offsets of $1.1 \sigma_{\rm point}$ and $1.6 \sigma_{\rm point}$ for 100 \um\ and 160 \um, respectively. As such, we continue with our analysis assuming the third scenario is correct.

\section{Stable orbits and disc sizes}

Multiple star systems give rise to interesting dynamics. In this system we can expect stable orbits around each of the stars, circumbinary orbits around the two secondary stars, and circumtertiary orbits around the whole system. By using the semi-analytical expression based on simulations by \citet{holman1999} we can estimate the sizes of stable regions, expressed as critical semi-major axes, around each star in a binary (see also \citealt{wiegert2014} and references therein).

\begin{figure*}
    \begin{center}
        \includegraphics[width=150mm]{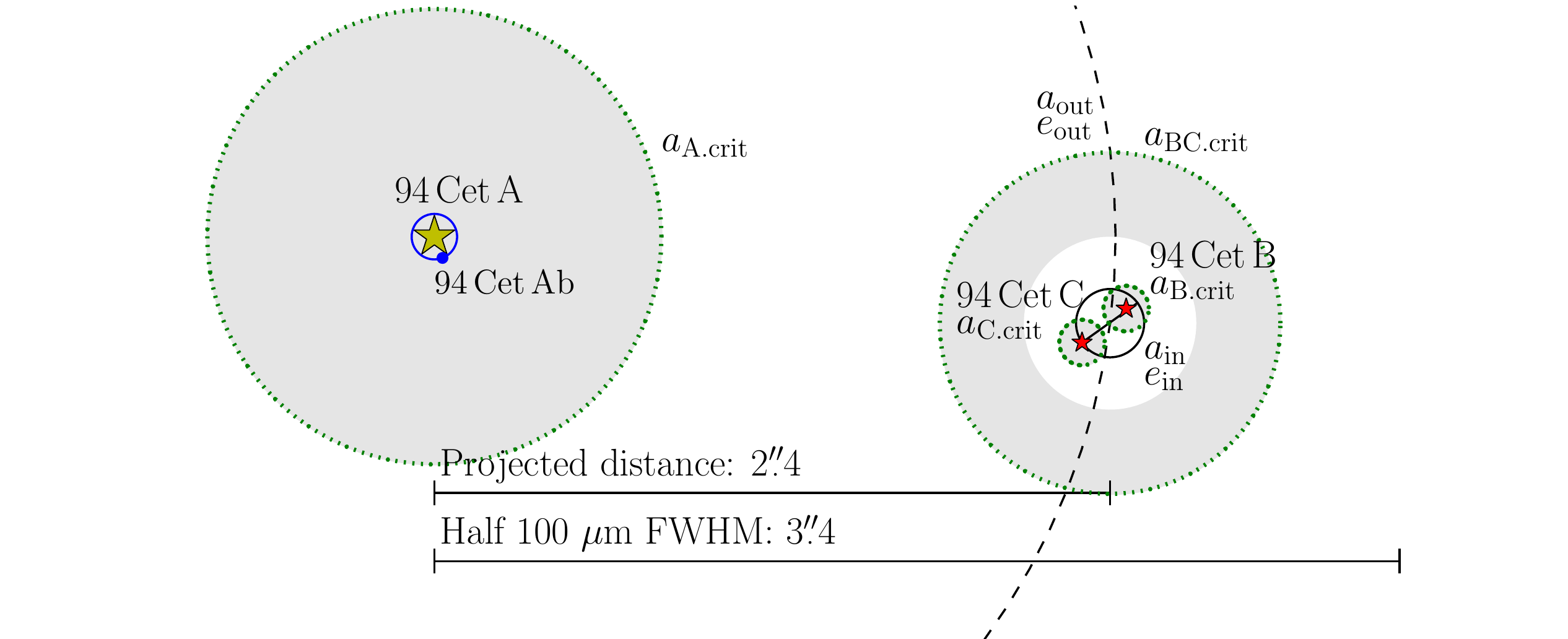}
        \caption{Schematic overview of the known components of the 94\,Cet system (not to scale). The \textit{outer orbit} denotes the orbit of the binary pair, 94\,Cet\,BC, around 94\,Cet\,A, where $a_{\rm out}$ is the semi-major axis (220\,AU), and $e_{\rm out}$ is the eccentricity (0.26). We denote the orbit of 94\,Cet\,C around 94\,Cet\,B as the \textit{inner orbit} and similarily are $a_{\rm in}$ and $e_{\rm in}$ the inner semi-major axis (0.99\,AU) and eccentricity (0.36). The blue dot and line indicates the orbit of the planet 94\,Cet\,Ab with a semi-major axis of 1.42\,AU, and eccentricity of 0.30. The green dotted lines are the critical semi-major axes and where we can expect stable orbits are shown as shaded regions. These are denoted as $a_{\rm A.crit}$, $a_{\rm B.crit}$, $a_{\rm C.crit}$, and $a_{\rm BC.crit}$ ($a_{\rm ABC.crit}$ is outside this figure). The projected distance refers to the angular distance between the stars during the observations and this is compared to half of the \textit{Herschel} beam FWHM at 100\um .}
        \label{hip14954schemorbit}
    \end{center}
\end{figure*}

We proceeded by calculating the critical semi-major axes for 94\,Cet\,B and C independently ($a_{\rm B.crit}$ and $a_{\rm C.crit}$), followed by coupling them and forming the BC binary component ($a_{\rm BC.crit}$). The BC component was then coupled with 94\,Cet\,A to form the A-(BC) tertiary system ($a_{\rm A.crit}$ and $a_{\rm ABC.crit}$, see Fig.\,\ref{hip14954schemorbit}). 

We found the following maximum radii for stable regions in the 94\,Cet system, with error estimates in parentheses, and the corresponding angular sizes based on the distance to the system,
\begin{equation}
\begin{cases}
a_{\rm A.crit} \approx 47.1\,(14\%)\,{\rm AU} = \asecdot{1}{8}\,(15\%) \\
a_{\rm B.crit} \approx     0.2\,(22\%)\,{\rm AU} = \asecdot{0}{009}\,(22\%) \\
a_{\rm C.crit} \approx     0.1\,(39\%)\,{\rm AU} = \asecdot{0}{004}\,(44\%) \\
a_{\rm BC.crit} \approx  36.4\,(22\%)\,{\rm AU} = \asecdot{1}{6}\,(23\%) .
\end{cases}
\label{acrits}
\end{equation}

The fifth critical semi-major axis, i.e. for circumtertiary orbits, can be approximated to be at least three times the semi-major axis \citep{wiegert1997,holman1999} which corresponds to $a_{\rm ABC.crit}\,\sim$\,660\,AU from the system barycentre. The western source is at an angular distance of \asecdot{35}{1} which corresponds to 794\,AU at this distance. Its position angle and distance fits well with a possible circumtertiary dust ring.

Although this disc would extend on scales of several hundreds of AU, it is not excluded to have a collisionally active belt. Firstly, because significantly extended debris discs have already been observed around solar-like stars, for instance around HD\,202628 where the disc extends up to $\sim 200\,$AU \citep{krist2012}, HD\,207129 with a belt at $\sim 160\,$AU \citep{lohne2012}, or $\zeta^{2}$\,Ret (HIP\,15371, \citealt{faramaz2014}), which has a wide stellar binary companion and a resolved, asymmetric disc at $\sim 100$\,AU. Secondly, the presence of a massive gravitational perturber, such as the BC pair orbiting the primary, is expected to increase the eccentricities of planetesimals inside the belt and raise the collisional activity, all the more since it is on an eccentric orbit. 

Due to its large extent, this system may have suffered from perturbations from galactic tides and passing stars. Indeed, this type of perturbations is expected to induce pseudo-random losses of angular momentum on bodies orbiting their host star at large separations, whether these bodies are planetesimals \citep{heisler1986} or stellar binary companions \citep{kaib2013}. Consequently, the orbits of these bodies tend to acquire small pericentres, which, in the case of planetesimals, will potentially lead to cometary activity \citep{kaib2009}. On the other hand, if it is a stellar binary companion, the survival of material around the primary may be more endangered and the architecture of any formed planetary system may be dramatically reshaped \citep{kaib2013}. These effects are expected to be significant for old (Gyr) systems and for bodies orbiting at very large separations (between $10^3$ and $10^4$\,AU). Since the separations involved in the case of the 94\,Cet system are smaller than these values, this system is most probably not affected significantly by galactic tides and passing stars.

As mentioned before, \citet{eiroa2013}, using a black-body fit, attempted to retrieve the dust location. They found a disc radius of 95\,AU, far inside the unstable region. Using the comparisons between black-body radii and resolved radii by \citet{pawellek2015} we finally estimate a range of \lq true\rq\ radii for this black body radius, i.e. 267 -- 455\,AU when taking error bars into account.

\section{Numerical study}

The resulting \lq true\rq\ radii are not sufficiently accurate and therefore we resort to full radiative transfer modelling to derive more accurate values. We also use N-body symplectic simulations as a complementary approach to our analytical estimates of the locations of stable orbits in this system.

\subsection{Radiative transfer modelling}

We use the Monte-Carlo based program {\sc radmc-3d} presented in \citet{dullemond2012}\footnote{\url{http://www.ita.uni-heidelberg.de/~dullemond/software/radmc-3d/}}. We use the disc constraints previously derived and combine them with {\sc radmc-3d} to obtain simulated SEDs and images of the different disc models. The parameters we used are summarized in Table\,\ref{dustproperties} and described hereafter.

\subsubsection{Dust properties}

\begin{table}
    \caption{Parameters for {\sc radmc-3d} simulations.}
    \label{dustproperties}
    \begin{center}
    \begin{tabular}{lc}
\hline\hline
    Parameter & Value \\
\hline
    Grain size range, $a$ (\um )                                       & 8 to $10^3$$^*$ \\
    Blow out radius, $a_{\rm blow-out}$ (\um )                         & 1.3$^a$ \\
    Grain density, $\rho$ (g\,cm$^{-3}$)                               & 2.5 \\
    Size distr., $n(a) \propto a^{-q}$ (cm$^{-3}$)                     & $q = 2$ to 4$^b$ \\
    Surface density, $\Sigma (r) \propto r^{-\gamma}$ (g\, cm$^{-2}$)  & $\gamma = -2$ to $2$ \\
    Vertical distribution, $h(r)$ (AU)                                 & $0.1 \times r$$^c$ \\
    Absorption coeff., $\kappa_{\rm abs}(a,q,\rho)$ (cm$^2$\,g$^{-1}$) & $\kappa_{\rm ext}\,(1 - \eta)$$^{d,e}$ \\
    Scatter coeff., $\kappa_{\rm scat}(a,q,\rho)$ (cm$^2$\,g$^{-1}$)   & $\kappa_{\rm ext} \eta$$^{d,e}$ \\
    Inner disc radius, $r_{\rm in}$ (AU)                               & $r(T_{\rm vap})$$^\dagger$ \\
    Outer disc radius, $r_{\rm out}$ (AU)                              & $a_{\rm crit}$$^\ddagger$ \\
\hline
    \end{tabular}
    \end{center}
    \begin{list}{}{}
    \item[References:]
            $(a)$ \citet{plavchan2009}.
            $(b)$ \citet{dohnanyi1969}.
            $(c)$ \citet{artymowicz1997}.
            $(d)$ \citet{miyake1993}.
            $(e)$ \citet{inoue2008}.
    \item[$*$:] $a_{\rm min} \approx 6\times a_{\rm blow-out}$ \citep{wyatt2011,lohne2012,thebault2016}.\\
$\dagger$: Inner disc radius is limited by vaporisation temperature, $T_{\rm vap} = $ 1800 to 1300\,K \citep{pollack1994,moromartin2013}.\\
$\ddagger$: See Equation\,\ref{acrits} and dynamical simulations.
    \end{list}
\end{table}

The mass absorption coefficient, $\kappa_{\rm abs}$, describes how well the dust grains absorb and re-emit radiation. We obtained it from the extinction coefficients ($\kappa_{\rm ext}$) of \citet{miyake1993}, who studied the effects of different grain sizes and size distributions of compact spherical silicate grains. The minimum grain size, that dominates the emission, is assumed to be $\sim \lambda / (2 \pi)$. We will be using the 100 \um\ data because of its better resolution and can assume minimum observed sizes between 10 and 100\,\um . The extinction coefficients are normalised with a gas-to-dust ratio of 100 (see \citealt{liseau2015}, and references therein). 

We may compare these opacities with other more recent studies: \citet{weingartner2001}, \citet{zubko2004} and \citet{draine2006} for instellar dust, and \citet{kataoka2014} for dust aggregates in protoplanetary discs. All these works show mass absorption coefficients between 1 and 10\,cm$^2$\,g$^{-1}$ at wavelengths around 300\,\um . However, they assume smaller grains than we expect in a circumstellar environment. Grains smaller than 10\,\um\ tend to exhibit up to two orders of magnitude higher absorption at shorter wavelengths ($<\,100$\,\um ) and stronger silicon features than grains of, e.g., 1\,mm. The final grain size range and size distribution exponent, $q$, are also significant contributors to the inferred total dust disc mass. $q = 3.5$ was first suggested by \citet{dohnanyi1969} but we also vary it between 2 and 4 in steps of 0.5.

The lower size limit is based on an inferred blow-out radius of the grains ($a_{\rm blow-out}$), i.e., the smallest possible grains allowed due to stellar pressure forces \citep{plavchan2009,wiegert2014}. Simulations \citep{wyatt2011,lohne2012,thebault2016} show that the lower cut-off is smooth and a good approximation of the smallest allowed grain size should be around 6 times the blow-out radius. We can compute the blow-out radius by assuming a mass grain density of 2.5\,g\,cm$^{-3}$ and obtain 1.3\,\um\ for 94\,Cet\,A. \citet{kirchschlager2013} found that only stars with effective temperatures higher than 5250 K have a determined blow-out radius and, since the companion stars are of M-type with effective temperatures under 4000 K, it is not expected for them to have a blow-out radius. Nevertheless, we assume the much brighter primary star generates a blow-out radius for the complete system. Since the larger grains do not contribute to the thermal emission of the dust, we fix the upper grain size to 1\,mm; the same value as several studies which allows us to have comparable results.

The albedo ($\eta$) may vary widely depending on grain constituents and possible ice covering. \citet{miyake1993} and \citet{inoue2008} both studied the albedo of grains with different sizes and with/without ice covering. They show that the albedo for small silicate particles ($a \lesssim$ 10\,\um ) is relatively stable between 0.5 and 0.6 and then drops down to zero at longer wavelengths between 100 to 500\,\um .

We estimate the optical depth at three different wavelengths: 0.5, 100 and 160\,\um . We use the chosen extinction coefficients and cloud densities of $10^{-20}$ -- $10^{-19}$\,g\,cm$^{-3}$ to obtain values of $0.3$ -- $3 \times 10^{-2}$ at optical wavelengths and $0.6$ -- $1 \times 10^{-3}$ at the far-infrared (FIR) wavelengths. From these values we can assume that the discs are optically thin.

We use {\sc radmc-3d}, which simulates the whole wavelength range and takes scattering into account, to simulate the heating processes at optical wavelengths, where the disc is less optically thin than at FIR. With more extreme disc models (e.g. higher densities) it might even be optical thick, and we would risk overestimating the emission at FIR and underestimate the mass if we were not using radiative transfer simulations.

Similarly as in \citet{wiegert2014} we start the simulations with a \lq standard\rq\ disc using the grain size power-law distribution of index $q = 3.5$ (homogeneously distributed in the disc) and radial disc surface density $\gamma = 1$. Keep in mind that a $\gamma\,<\,-2$ is in reality a ring at the outer edge of the allowed radius, and a $\gamma\,>\,2$ is instead a dust ring just outside the vaporisation radius. There were only small differences in the resulting SEDs with $\gamma$ outside the range of $-2$ to 2.

The accuracy of each model was inspected by eye and quantified with a reduced $\chi^2$ expressed as $1/N\,\times\,\sum_{\nu }\,\left[\,(S_\nu^{\rm obs}\,-\,S_\nu^{\rm model}\,)\,/\sigma\,\right]^2$, where $S_\nu^{\rm obs}$ is the observed flux density, $S_\nu^{\rm model}$ is the corresponding model flux density, and $N$ is the number of data points computed for each wavelength, that is, at 70, 100 and 160\,\um , as this is where the emission was found. Through an iterative process we explored the parameter space to test the degeneracy of each model.

The dust emission models are quantified by the luminosity ratio between the stellar total luminosity and the dust emission luminosity, i.e. $f_{\rm d} \equiv L_{\rm dust} / L_\star$.

\begin{figure}
    \begin{center}
        \includegraphics[width=75mm]{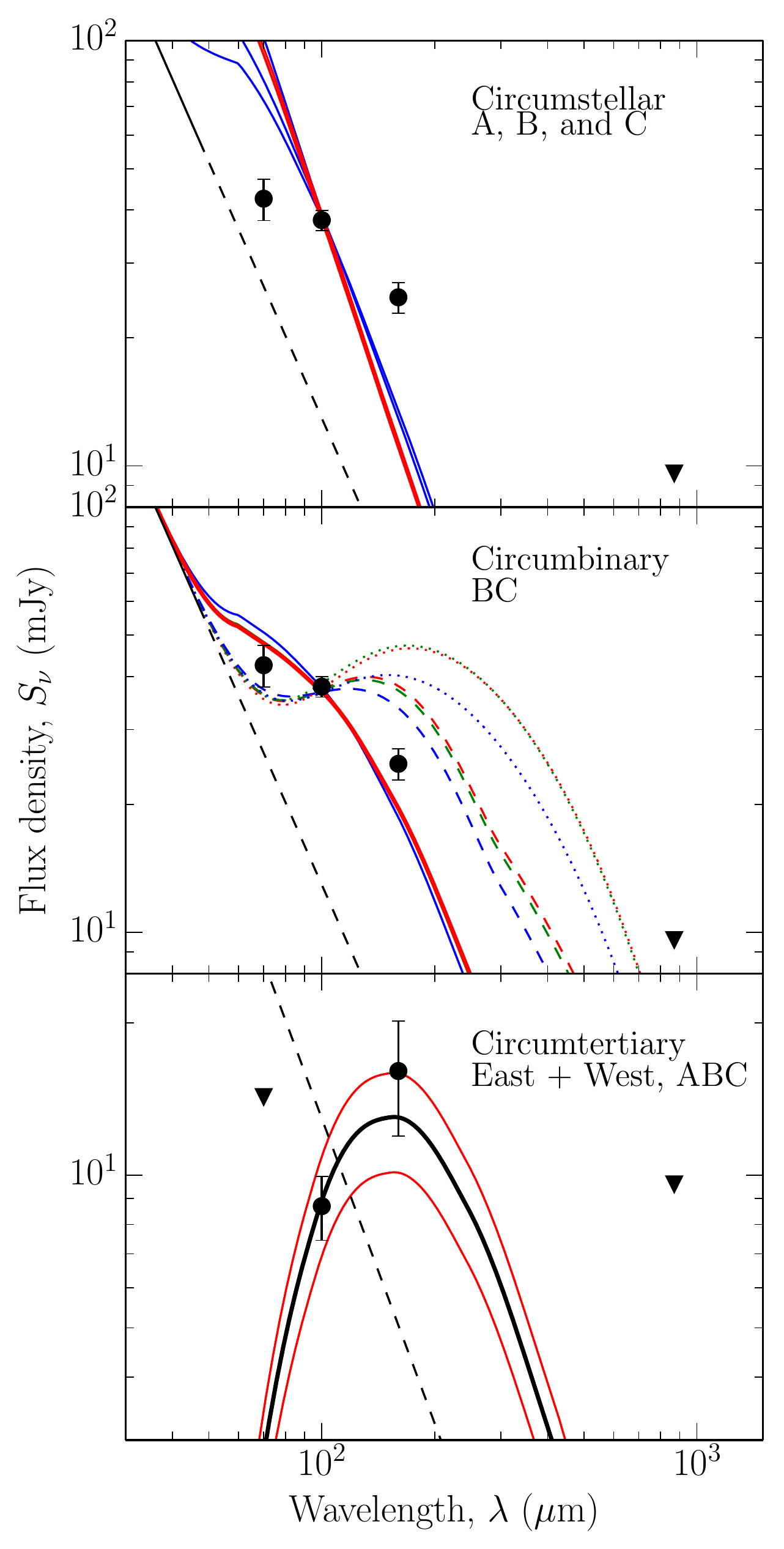}
        \caption{94\,Cet excess models. The thin black line is the model stellar photosphere and the dashed line indicates where a Rayleigh-Jeans tail is assumed. Black dots are photometric data points and the black triangle is a 3\,$\sigma$ upper limit. Top panel shows circumstellar discs, both circumprimary (blue) and circumsecondary (red). Middle panel shows circumbinary discs around the secondary pair of stars where the best fit is shown with a thick red line. Models with $\gamma = -2$ are red, $\gamma = 0$ are green, and $\gamma = 2$ are blue, and $q=3.5$ with solid lines, $q=3.0$ with dashed lines, and $q=2.5$ with dotted lines. The lower panel shows the best fit circumtertiary dust (black curve) with error estimates as thin red curves, and the data points are the combined flux density of the eastern and western extensions.}
        \label{hip14954sedmodels}
    \end{center}
\end{figure}

\subsubsection{Circumstellar disc SEDs}

The results of radiative transfer simulations of circumstellar discs are shown in the top panel of Fig.\,\,\ref{hip14954sedmodels}. The circumprimary disc and companion star circumstellar discs were simulated separately but are shown together. The models were normalised at the 100\,\um\ flux density.

The radius of the circumprimary disc may be located at up to 54\,AU. The PACS beam radius at the stellar distance corresponds to 81 and 130\,AU for 100 and 160\,\um , respectively, which implies that the disc would not be resolved. The inner radius was set to 0.1\,AU. There is a planet at 1.4\,AU that could form gaps in such a disc, however the available data is insufficient for this to be visible and the models presented here do not include these effects.

The simulated circumprimary disc is in general too warm to fit the data (average temperatures are around $60 \pm 20$\,K). Rings at the outermost edge of the stable region were too warm.

In the case of the companion pair BC, the small orbit significantly limits the radii of circumstellar discs to less than 0.2\,AU. The simulated SEDs in the top panel of Fig.\,\ref{hip14954sedmodels} (red curves) are based on discs with inner radius of 0.008\,AU and outer radii of 0.21 and 0.16\,AU for 94\,Cet\,B and C, respectively. The circumsecondary dust is too warm to fit the observations, with average dust temperature of $160\,\pm 70$\,K.

\subsubsection{Circumbinary disc SEDs}

The circumbinary disc refers to dust orbiting both the secondary stars, 94\,Cet\,B and C. The radial limits from the companion barycentre are 3\,AU to 40\,AU (\asecdot{0}{1} to \asecdot{1}{8}, Equation\,\ref{acrits}).

The resulting SEDs from {\sc radmc-3d} are shown in the middle panel of Fig.\,\ref{hip14954sedmodels}. Those with $q=3.5$ were too warm. Reducing the disc into a ring in the outermost parts of the stable region did not reduce the temperature significantly as this did not increase the average distance to the primary star. However, reducing the number of small grains and increasing the number of large grains, i.e. reducing the size distribution $q$ significantly cools the disc. Thus we also show simulated SEDs with $q = 3.0$ and 2.5 in the same figure.

The best fit dust model ($\chi^2 = 2.6$) corresponds to a disc with $\gamma = -2.0$, $q=3.5$, average dust temperature of $30.3 \pm 7.4$\,K, and a luminosity ratio of $f_{\rm d} = 4.6\pm 0.4\,\times 10^{-6}$. The inferred total dust mass would be $6.0\pm 0.5\,\times 10^{-2}\,M_{\rm Moon}$. Error estimates are based on the average error of 7.9\,per\,cent of the flux densities at 70, 100, and 160\,\um .

A formally better fit could probably be found with a $q$ slightly less than 3.5. However, we have too few SED measurements to constrain this better and decrease the $\chi^2$.

\subsubsection{Circumtertiary disc SEDs}

\begin{figure*}
    \begin{center}
        \includegraphics[width=150mm]{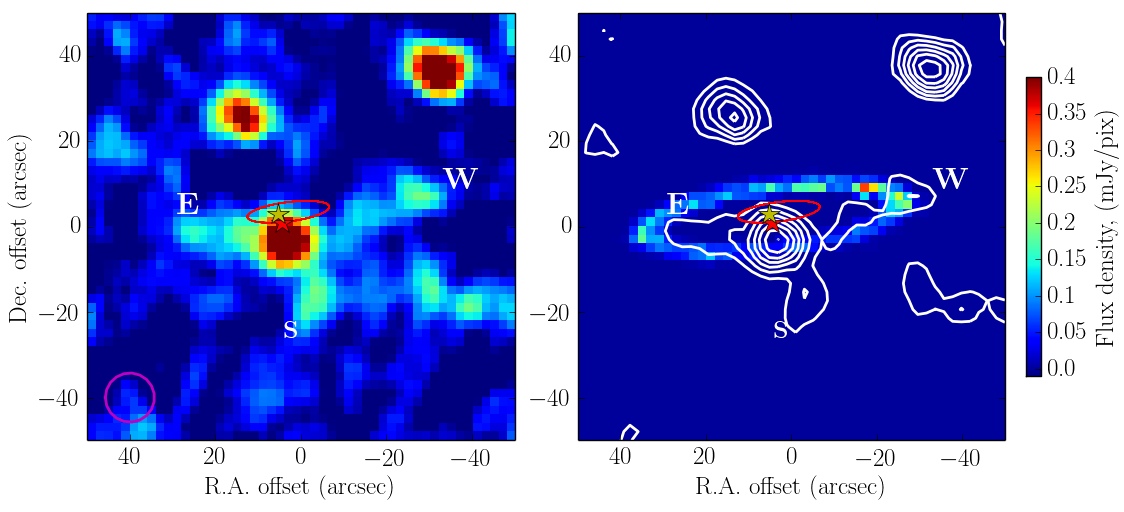}
        \caption{Comparison of observations with circumtertiary ring simulations. Left panel shows the observations at 160\,\um\ with the eastern (E), southern (S), and western (W) extensions, the real position of 94\,Cet (yellow and red stars), and the companion pair orbit around the primary (red ellipse). Magenta ring is the observation beam size. Right panel shows the simulated emission of a simulated circumtertiary ring and the 160\,\um\ observation is plotted on top as white contours. The east extension has flux densities $4.2 \pm 0.5$\,mJy at 100\,\um\ and $8.5 \pm 1.8$\,mJy at 160\,\um\ and the west extension has $4.5 \pm 0.7$\,mJy at 100\,\um\ and $7.6 \pm 2.3$\,mJy at 160\,\um .}
        \label{hip14954circumtren}
    \end{center}
\end{figure*}

The eastern and western extensions appear where a circumtertiary ring is expected. Fig.\,\ref{hip14954circumtren} shows a comparison between the observed emission and photometry from a simulated circumtertiary disc at 160\,\um\ based on symplectic particle simulations (see Section\,6.2).

We estimate the circumtertiary flux density from the combined fluxes of the eastern and western extensions to $8.7 \pm 1.3$\,mJy at 100\,\um\ and $16.1 \pm 4.1$\,mJy at 160\,\um\ as plotted in the lower panel of Fig.\,\ref{hip14954sedmodels}.

Initially the combined east plus west flux density appear to follow the black body of dust ring with a temperature corresponding to the inner dynamical stable radius, 660\,AU. The emission is however unconstrained on the Rayleigh-Jeans part of the SED and a \lq standard\rq\ dust model with $q=3.5$ and $\gamma=1.0$ is a sufficient input to {\sc radmc-3d}.

The best fit model corresponds to a dust ring with fractional luminosity $f_{\rm d} \approx 1.4\pm 0.3\,\times 10^{-6}$ and average temperature of $17.0 \pm 6.1$\,K. The temperature is unconstrained though and an upper limit of 30\,K is a better estimate. With $q=3.5$ the total dust mass would be $\sim 0.29 \pm 0.06\,M_{\rm Moon}$.

A possible upper limit of the ring radius could however be measured directly from the maximum distance between 94\,Cet and the outer part of W, i.e. $\sim 38$\arcsec\ or 860\,AU.

In Fig.\,\ref{hip14954circumtren} we see that a ring would exhibit peak flux densities at the outer edges to the west and east. This is also where we see the western emission, however at the eastern edge there is no detectable emission. Furthermore, the eastern extension corresponds better with where the gap would be under the assumption that the ring is in the same orbital plane as the stars. However, an edge-on ring would emit along that region and a very clumpy ring could possibly be able to fit.

Another explanation could come from the pericentre glow effect. Pericentre glow appears when a circumstellar ring is eccentric so that the pericentre is more heated than the other parts of the ring, resulting in a horseshoe shaped source. This effect may occur due to the influence of a companion as, e.g., a planet (\citealt{wyatt1999}, compare with the Fomalhaut ring, \citealt{acke2012}).

\subsection{N-body simulations}

Symplectic integration techniques take advantage of the fact that in a planetary system, the mass of the central body is much larger than all the other ones; however, they fail if all massive bodies have comparable masses, such as in multiple stellar systems. Therefore, we used the symplectic integrator HJS\footnote{Available at \url{http://ipag.osug.fr/~beusth/hjs.html}} (Hierarchical Jacobi Symplectic) of \citet{beust2003}, that allows the study of the dynamics of hierarchical stellar systems, provided that the hierarchical structure of the system is preserved along the integration. This code permits us to study the distribution of stable orbits for circumstellar material in this triple star system.

The initial conditions are summarized in Table\,\ref{hip14954dynsimtab}. All simulations contain 10,000 test particles with semi-major axes uniformly and randomly distributed between the initial inner and outer values. Their eccentricities are randomly distributed between 0 and 0.05, in order to mimic the low eccentricity-orbits one can expect at the end of the protoplanetary phase, as well as low inclinations randomly distributed between -3\adeg\ and 3\adeg\ relative the outer orbital plane. The remaining initial angles, longitude of ascending node, longitude of periastron and mean anomaly, are randomly distributed between 0 and $2\,\pi$. We use values corresponding to a cold disc, which is not necessarily true given the age of the system, however, we wish here to test our analytical predictions on the location of stable orbits in the current configuration of the system, so that this simple approach should be sufficient for our purpose.

\begin{table*}
    \caption{Particle disc simulations, initial and final conditions.}
    \label{hip14954dynsimtab}
    \begin{center}
    \begin{tabular}{llll}
\hline\hline
    Parameter                          & Circumprimary & Circumbinary & Circumtertiary \\
\hline
    Central star(s)                    & 94\,Cet\,A    & 94\,Cet\,BC  & 94\,Cet\,ABC \\
    Initial outer semi-major axis (AU) & 60            & 50           & 750 \\
    Final outer semi-major axis (AU)   & 50            & 40           & 770 \\
    Initial inner semi-major axis (AU) & 10            & 3            & 250 \\
    Final inner semi-major axis (AU)   & 10            & 2.5          & 590 \\
    Number of particles                & $10^4$        & $10^4$       & $10^4$ \\
    Length of simulation (Myr)         & 20            & 20           & 20 \\
\hline
    \end{tabular}
    \end{center}
\end{table*}

\begin{figure*}
    \begin{center}
        \includegraphics[width=170mm]{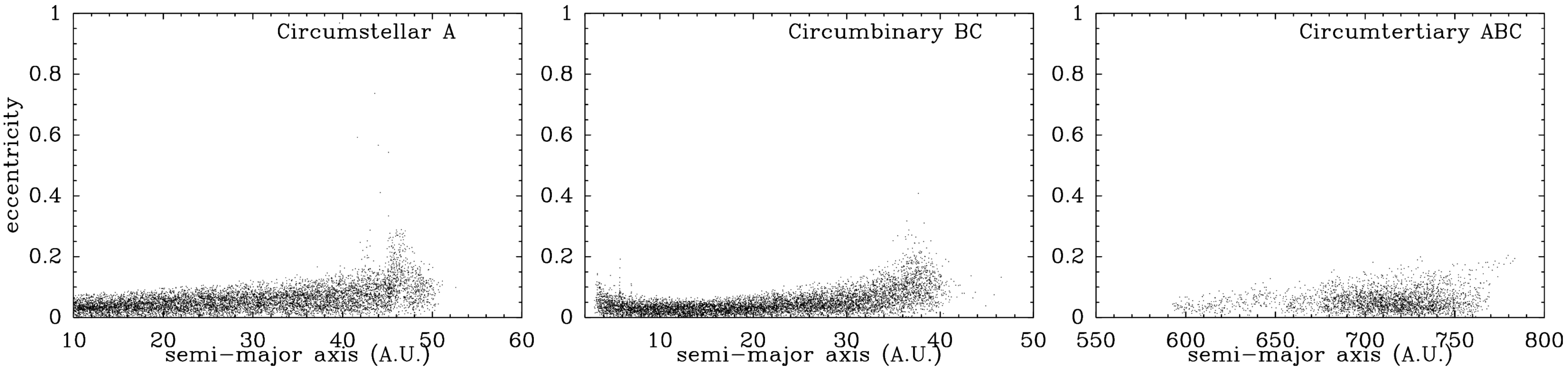}
        \caption{Symplectic particle simulation results. Final eccentricity of each particle orbit is shown against their semi-major axis for circumprimary, circumbinary, and circumtertiary discs respectively. Vertical structures appear on the first two panels. The enhancement of the eccentricities of the test-particles is due to the orbital eccentricity of the perturbers they are subject to where the disc is truncated. Indeed, in the case of the circumprimary disc, which is under the influence of the pair BC on an eccentric orbit, the enhancement appears at the outer edge of the disc only, while it appears both at the inner and outer edges in the cicumbinary case, since the disc is under the influence of the eccentric orbit of C around B at the inner edge, and under the influence of the eccentric orbit of a at the outer edge.}
        \label{hip14954dynsims}
    \end{center}
\end{figure*}

We ran three separate simulations, circumprimary (around 94\,Cet\,A), circumbinary (around 94\,Cet\,B and C), and circumtertiary. In each case, a timestep of 1/20 of the smallest orbital period was used, which ensures a conservation of energy with a typical error of $\sim 10^{-6}$ \citep{beust2003}. The results are shown in Fig.\,\ref{hip14954dynsims}. With a 20\,Myr simulation length ($\sim 1/100$ of the age of the system), that corresponds to 10,000 orbits of the secondary around the primary, we can assume that the system is stable.

The results are consistent with the estimates in Equation\,\ref{acrits} (and the discs used for {\sc radmc-3d}). The low density component between 600 and 670\,AU probably consists of particles in the process of being ejected from the system.

In conclusion, the semi-analytical expression from \citet{holman1999} is useful also for triple systems where the component's semi-major axes differs enough so that one can approximate the system as two separate binaries.

\section{Conclusions}

Multi-component systems pose many difficult questions concerning planet formation processes. 94\,Cet provides one of few unique opportunities of a case study of a triple stellar system \citep{roll2011a,roll2012} with at least one planet \citep{eggenberger2003} and FIR excess \citep{trilling2008,eiroa2013}. For these reasons and a marginally significant offset in the PACS data we aimed to model the stellar excess and simulate the orbital dynamics of the system.

\begin{itemize}
    \item[--] The central source fits well with that of a circumbinary disc around the companion pair (inside the dynamically stable radii) at $30.3 \pm 7.4$\,K, with a dust grain size distribution $q$ between 3 and 3.5, and fractional luminosity $f_{\rm d} = 4.6\pm 0.4\,\times 10^{-6}$. The disc extends from 3\,AU to 40\,AU from the companion stars' barycentre with a surface density distribution that was $\propto r^{2}$. Assuming $q=3.5$ and the grain size range of 8\,\um\ -- 1\,mm this corresponds to $6.0\pm 0.5\,\times 10^{-2}\,M_{\rm Moon}$.

    \item[--] The eastern and western extensions have the flux densities $8.7 \pm 1.3$\,mJy at 100\,\um\ and $16.1 \pm 4.1$\,mJy at 160\,\um , which corresponds well with the possibility of a circumtertiary ring of dust ($q=3.5$) with the fractional luminosity $f_{\rm d} \approx 1.4\pm 0.3\,\times 10^{-6}$ and temperature of $< 30$\,K. However the uncertainty is significant and the temperature is unconstrained. The ring was assumed to extend from an inner edge at 600 -- 650\,AU to $\sim 750$\,AU, and the initial surface density distribution of $\propto r^{-1}$.

    \item[--] The system and disc configurations are stable after 20 Myr. The particle discs were simulated with the symplectic integrator HJS \citep{beust2003} using $10^4$ particles for each disc.

\end{itemize}

Our models provide evidence for the possibility of both circumsecondary and circumteriary dust. It is possible that the lack of circumprimary (hotter) dust emission is due to additonal planets emptying the circumprimary neighbourhood and our evidence for the circumtertiary ring is only tentative. As such it would be useful to further constrain the nature of this system through additional observations. 

Even without \textit{Herschel} there are a few possibilities available. ALMA, for example, could probably not observe any circumtertiary dust, but may be able to confirm our findings on the circumsecondary dust, and the LMT (Large Millimeter Telescope) in Mexico can observe 94\,Cet during autumn and could reach a $3\,\sigma$ of 0.3\,mJy in 36\,hours, possibly enough to reach the dust emission.

\section*{Acknowledgements}

We would like to thank R.\,Liseau, and also C.\,Eiroa, G.\,Kennedy, A.\,Krivov, J.P.\,Marshall, and K.\,Torstensson for their many comments, insights, and support for this project. We appreciate the continued support of the Swedish National Space Board (SNSB) for our {\it Herschel} projects. V.\,Faramaz acknowledges the support from FONDECYT Postdoctoral Fellowships, project no 3150106, and the support from the Millenium Nucleus RC130007 (Chilean Ministry of Economy). F. Cruz-Saenz de Miera is supported by CONACyT research grant SEP-2011-169554. Many thanks to Herv\'e Beust for his guidance using the HJS code.

This research has made use of the Exoplanet Orbit Database and the Exoplanet Data Explorer at \url{exoplanets.org}, the SIMBAD database, operated at CDS, Strasbourg, France, the Two Micron All Sky Survey, which is a joint project of the University of Massachusetts and the Infrared Processing and Analysis Center/California Institute of Technology, funded by the National Aeronautics and Space Administration and the National Science Foundation, and the NASA/IPAC Extragalactic Database (NED) which is operated by the Jet Propulsion Laboratory, California Institute of Technology, under contract with the National Aeronautics and Space Administration.

% ---------------------------------------------------------------------------- %
% references
%
\bibliographystyle{mnras}
\bibliography{wiegert94cet}
%
% ---------------------------------------------------------------------------- %

\appendix

\section{Background galaxies}

We list background sources found in the PACS fields and LABOCA field in Tables\,\ref{spectralindices} and \ref{labocafluxes}.

Five background sources with S/N\,$> 3$ were found in the PACS field. For the positions of the sources 94\,Cet-2 and 5, we found the catalogued sources SSTSL2 J031244.01-011111.8 and SSTSL2 J031248.04-010850.9, respectively, in the \textit{Spitzer} Heritage Archive\footnote{\label{ipacurl}Searched through the NASA/IPAC Infrared Science Archive, Caltech/JPL. \url{https://irsa.ipac.caltech.edu/}}. However, they were not found in other large catalogues (e.g. Two Micron All Sky Survey$^{\ref{ipacurl}}$, \citealt{skrutskie2006}, or NASA/IPAC Extragalactic Database\footnote{The NASA/IPAC Extragalactic Database (NED) is operated by the Jet Propulsion Laboratory, California Institute of Technology, under contract with the National Aeronautics and Space Administration. \url{http://ned.ipac.caltech.edu/}}). They were not visible in the Palomar Sky Survey, possibly due to the large PSF of the star in infrared. However, in the MIPS 24\,\um\ data we notice weak sources (i.e. S/N\,$< 3$) at the positions of the PACS background sources, and that 94\,Cet-4 coincides with a ring feature (a beam artefact) that appears around the source at 24\,\um .

The field appears to lack previously detected sources, there is however a great number of sources to the west of the field. The lack of catalogued sources is not surprising though as the stars in the DUNES catalogue were deliberately chosen to avoid regions of noisy background as e.g. the galactic plane. In the POSS infrared images, 94\,Cet-3 possibly appears, however very weakly. The other sources are either covered by the large PSF from the star or too faint. We do see the sources denoted as 94\,Cet-2, 3, and 4 in the MIPS 70\,\um\ data, however, all of these have S/N\,$<\,3$. And as mentioned, the stellar proper motion between \textit{Spitzer} and \textit{Herschel} observations are just slightly more than 1\,arcsec.

\begin{table}
    \caption{PACS-field background sources, coordinates, and spectral indices with wavelengths 160\,\um\ and 870\,\um .}
    \label{spectralindices}
    \begin{center}
    \begin{tabular}{lccl}

\hline\hline
    Source & R.A. (J2000) & Dec. (J2000)         & Spectral \\
           & h m s        & \adeg\ \amin\ \asec\ & index $\alpha^*$ \\
\hline
    94\,Cet$^\dagger$    & 3 12 46.43 &  -1 11 51.8 & $\le 0.56$ \\
    94\,Cet-1            & 3 12 41.56 &  -1 12 12.6 & $\le 0.51$ \\
    94\,Cet-2$^\ddagger$ & 3 12 44.02 &  -1 11 12.5 & $\le 0.64$ \\
    94\,Cet-3            & 3 12 44.08 &  -1 10 33.5 & $\le 0.43$ \\
    94\,Cet-4            & 3 12 47.08 &  -1 11 22.8 & $\le 0.38$ \\
    94\,Cet-5$^\sharp$   & 3 12 48.05 &  -1 08 51.1 & $\le 0.77$ \\
    Average index        &            &&\multirow{2}{*}{$\le 0.64$}\\
    of background        &            &             & \\
\hline
    \end{tabular}
    \end{center}
    \begin{list}{}{}
    \item[Comments.] $^*$ Spectral index for $\lambda = 870$ and 160\,\um , and defined as $S_\nu \propto \nu^\alpha$.$^\dagger$ Observed coordinates, not from references.$^\ddagger$ Coordinates correspond with SSTSL2 J031244.01-011111.8.$^\sharp$ Coordinates correspond with SSTSL2 J031248.04-010850.9.
    \end{list}
\end{table}

\begin{table}
    \caption{LABOCA-field background sources, coordinates, and flux densities.}
    \label{labocafluxes}
    \begin{center}
    \begin{tabular}{lccl}
\hline\hline
    Source & R.A. (J2000) & Dec. (J2000)         & $S_\nu$ (mJy) \\
           & h m s        & \adeg\ \amin\ \asec\ &               \\
\hline
    94\,Cet-L1      & 3 12 39.08 &  -1 17 06.8 & $18.70 \pm 5.93$ \\
    94\,Cet-L2      & 3 12 42.30 &  -1 15 32.2 & $14.55 \pm 3.73$ \\
    94\,Cet-L3      & 3 12 46.02 &  -1 16 16.3 & $16.13 \pm 4.81$ \\
    94\,Cet-L4      & 3 12 46.56 &  -1 07 00.0 & $16.56 \pm 5.49$ \\
    94\,Cet-L5      & 3 12 47.37 &  -1 13 20.3 & $14.62 \pm 2.73$ \\
    94\,Cet-L6      & 3 12 51.89 &  -1 09 36.0 & $17.97 \pm 4.11$ \\
    94\,Cet-L7      & 3 12 58.34 &  -1 13 02.5 & $16.05 \pm 3.90$ \\
\hline
    \end{tabular}
    \end{center}
\end{table}

\begin{figure}
    \begin{center}
        \includegraphics[width=80mm]{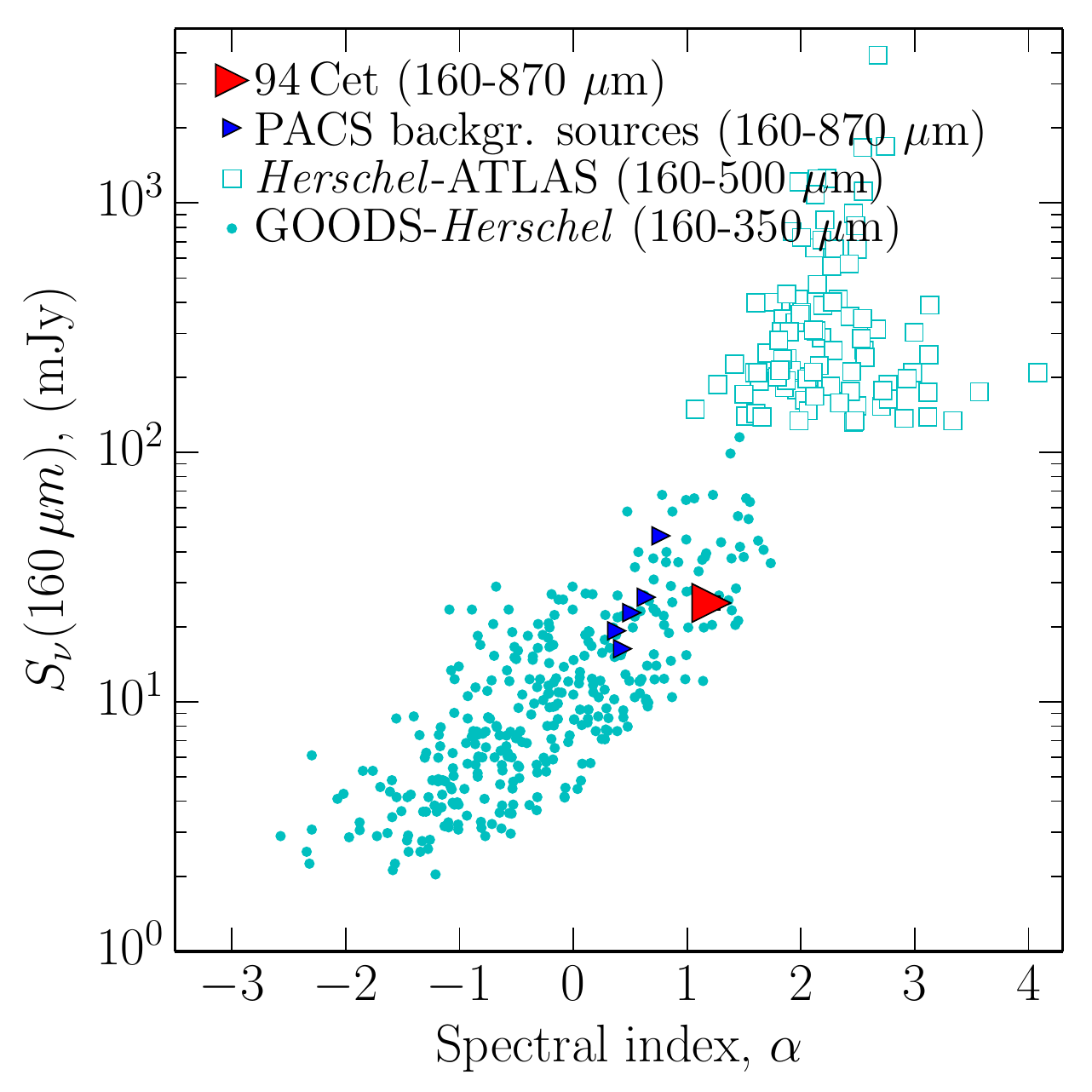}
        \caption{Spectral index map of the sources listed in Table\,\ref{spectralindices} and FIR galaxies. Cyan dots are spectral indices of FIR galaxies from the north field of the GOODS-\textit{Herschel} \citep{elbaz2011,elbaz2013} with $\lambda = 160$ to 350 and 500\,\um . Cyan squares are spectral indices of FIR galaxies from \textit{Herschel}-ATLAS \citep{rigby2011} with $\lambda = 160$ to 500\,\um . Red triangle indicate 94\,Cet upper limit spectral index (using $1\,\sigma$ upper limit at 870\,\um ), and the blue triangles are background sources upper limits with $\lambda = 160$ to 870\,\um\ ($3\,\sigma$ upper limit at 870\,\um ).}
        \label{colourmap}
    \end{center}
\end{figure}

In Table\,\ref{spectralindices} we clearly see that all sources have similar upper limit spectral indices (upper limits due to non-detections in the LABOCA data) as the central source. We also compare these data with larger samples of FIR galaxies from GOODS-\textit{Herschel} \citep{elbaz2011,elbaz2013} and \textit{Herschel}-ATLAS \citep{rigby2011} in Fig.\,\ref{colourmap}. We define the spectral index $\alpha$ as $S_\nu\,\propto\,\nu^{\, \alpha}$ with wavelengths from $\lambda = 160$ to 350, 500, or 870\,\um\ depending on available data.

However, we also see that it is difficult to distinguish the stellar excess from the background sources and other FIR galaxies by looking at the spectral indices. This is not strange as we are comparing extragalactic dust with circumstellar dust at these wavelengths.

The similarities are clearer if we compare with model SEDs of FIR galaxies (redshifts of $z \sim 0.2$ to 2) from \citet{chary2001}\footnote{Templates and instructions available at \url{http://david.elbaz3.free.fr/astro_codes/chary_elbaz.html}}. These show that FIR galaxies tend to have peak flux densities at PACS wavelengths while at optical and near infrared wavelengths the flux densities are significantly lower. If we compare with 2MASS completeness limits we see that not even the most extreme galaxies (with highest star forming rates) at redshifts higher than 0.5 should be detectable at optical or near infrared wavelengths.

Most of the PACS background sources are likely FIR galaxies. However, the central source is extended in nature and fits the positions and position angles of the stellar components, its orbit, and the expected size of a circumtertiary ring. With a 2\,per\,cent probability of coincidental alignment we find it very improbable that the dust emission seen at 94\,Cet could be attributed to a FIR background galaxy.

%
% End of document ----------------------------------------------------------- %
% Don't change these lines
\bsp	% typesetting comment
\label{lastpage}
\end{document}